# LIPOARABINOMANNAN-BASED TUBERCULOSIS DIAGNOSIS USING A FIBER CAVITY RING DOWN BIOSENSOR

UBAID ULLAH[1], SEERAT SALEEM[2], MUDDASSAR FAROOQ[3], BASIT YAMEEN[2,*], AND M. IMRAN CHEEMA[1,*]

[1]Department of Electrical Engineering, Lahore University of Management Sciences, Lahore (Pakistan)
[2]Department of Chemistry and Chemical Engineering, Lahore University of Management Sciences, Lahore (Pakistan)
[3]CureMD Healthcare, 80 Pine Street, 21st Floor | New York, NY 10005 | USA
[*]basit.yameen@lums.edu.pk; [*]imran.cheema@lums.edu.pk

**Abstract:**

Despite existing for millennia, tuberculosis (TB) remains a persistent global health challenge. A significant obstacle in controlling TB spread is the need for a rapid, portable, sensitive, and accurate diagnostic test. Currently, sputum culture stands as a benchmark test for TB diagnosis. Although highly reliable, it necessitates advanced laboratory facilities and involves considerable testing time. In this context, we present a rapid, portable, and cost-effective optical fiber sensor designed to measure lipoarabinomannan (LAM), a TB biomarker found in patients' urine samples. Our sensing approach is based on the applications of phase shift-cavity ringdown spectroscopy (PS-CRDS) to an optical fiber cavity created by two fiber Bragg gratings. A tapered fiber is spliced inside the optical cavity to serve as the sensing head. We functionalize the tapered fiber surface with anti-LAM antigen CS-35 through a unique chemistry, creating a strong affinity for LAM molecules. We measure the phase difference between the cavity transmission and the reference modulating signal at the cavity output. The measured phase is directly proportional to the injected LAM concentrations in aqueous solutions over the sensing head. Our demonstrated sensor provides a detection limit of 10 pg/mL and a sensitivity of 2.6°/ng/mL. This sensor holds promise for numerous applications in the healthcare sector, particularly in low-resource settings.

## 1. Introduction

Tuberculosis (TB) is one of the deadliest global health threats, causing over 1.4 million deaths and 10 million new cases yearly, with a disproportionate burden on low and middle-income countries [1]. Currently, sputum culture is the gold standard test for TB detection [2], but this method entails a protracted turnaround time of four to eight weeks. Moreover, it necessitates highly trained laboratory personnel and resource-intensive laboratory environments to execute multiple processing steps. Patients must endure multiple visits to diagnostic centers, presenting a substantial challenge, particularly in remote and low-resource regions.

One of the significant WHO objectives to fight TB is the development of a rapid, sensitive, specific, portable, and cost-effective diagnostic TB sensor [3]. In recent years, various optical and non-optical sensors have emerged for TB detection in human breath, blood, sputum, and urine [4–8]. For instance, nucleic acid amplification tests (NAATs) like GeneXpert and its variants employ patient sputum for rapid TB diagnosis. Nevertheless, these systems, though increasingly prevalent in TB diagnostics, still need to achieve WHO-recommended sensitivity, specificity, and cost-effectiveness standards [9]. In contrast, a Raman spectroscopy-based TB diagnostic system has recently demonstrated the requisite sensitivity, specificity, and portability, although cost remains a limiting factor [10].

Furthermore, there is a pressing need to monitor the disease state once treatment commences to optimize medications. Monitoring the treatment response in NAAT-based systems is challenging [9], while Raman-based systems, integrated with AI, have shown limited capabilities [10]. Therefore, time-consuming culture tests remain the primary method for assessing treatment response.

Sputum-based diagnostic systems face various challenges, including sample variations among patients and difficulties in obtaining sputum samples from children and elderly individuals. Consequently, non-sputum-based tests are highly sought, with researchers exploring potential TB biomarkers in blood, breath, and urine. Among these media, urine is appealing due to its easy collection from children and the elderly and a lower risk of infection during collection.

Lipoarabinomannan (LAM), a glycolipid found in the cell walls of TB-causing mycobacteria, has emerged as a promising urine-based TB biomarker, first demonstrated in the early 2000s [11]. Subsequently, researchers have developed various LAM detection assays for TB diagnosis. Notably, the AlereLAM test, while cost-effective, needs to improve in terms of desired sensitivity. However, the Fujifilm TB LAM test appears as a promising solution, particularly for children and individuals with HIV, aiming to enhance the sensitivity of AlereLAM. Additionally, innovative approaches, such as nanotechnology and GC/MS methods, show promise in augmenting LAM detection sensitivity. Nevertheless, challenges persist in achieving the desired levels of sensitivity, accuracy, and cost-effectiveness as recommended by the WHO [12,13].

One potential avenue toward developing a viable LAM detection test for point-of-care TB diagnosis involves using optical biosensors. However, research in the realm of LAM photonics biosensors remains limited. In a notable work, researchers employed a Mach-Zhender interferometer on a chip, coupled with a broadband source, spectral filters, and a CMOS camera, to establish a correlation between LAM concentration and spectral shifts. The work demonstrated a detection limit of 475 pg/mL [14].

In the present work, we demonstrate the first application of fiber cavity ring down spectroscopy for LAM detection toward TB diagnosis. We build a fiber cavity using fiber Bragg gratings at 1550 nm with a tapered fiber as a sensing head. The tapered fiber is functionalized with monoclonal antibodies employing a novel chemical functionalization protocol for LAM binding. Using the phase shift-cavity ring down spectroscopy (PS-CRDS) principle, we show the LAM detection limit of 10 pg/mL in aqueous solutions.

We now describe the rest of the paper. Section 2 describes various materials and methods in the work, including tapered fiber fabrication, chemical functionalization, and experimental setup. Section 3 provides results and related discussions, followed by concluding remarks in Section 4.

## 2. Materials and Methods

*2.1 Sensing Head*

We employ tapered fibers as sensing heads. Tapered fibers are engineered thin optical fibers that allow a portion of the light to interact with the surrounding analyte. We use a computer-controlled motorized stage to pull an SMF-28 fiber while heating it with a flame. This makes the fiber thinner. We attach the thin fiber to a U-shaped copper-clad printed circuit board with nichrome soldering. Then, we carefully take it out of the tapering setup. The soldering is necessary because glues or epoxies can be damaged by harsh chemicals like the piranha solution used in the surface functionalization steps [15].

*2.2 Surface Functionalization*

The various steps involved in surface functionalization are outlined in the schematic diagram presented in Fig. 1. The process begins by activating optical fibers using a piranha solution

($H_2SO_4$: $H_2O_2$ 7:3), which generates a reactive functional group on the surface. Subsequently, the fibers undergo treatment with APTES, facilitating the installation of amino groups on their surface. The next stage involves generating surface aldehydic groups through the use of glutaraldehyde. This compound reacts with the amino groups of the antibody, enabling the functionalization of the monoclonal anti-mycobacterium tuberculosis LAM on the fiber surface. Specifically, the monoclonal anti-mycobacterium tuberculosis LAM, clone CS-35 (NR-13811), can bind specifically with the LAM. In the following sections, we elaborate on our tapered fibers functionalization protocol details.

2.2.1 Materials

Concentrated sulphuric acid ($H_2SO_4$), hydrogen peroxide 30% ($H_2O_2$), acetone, deionized water, 3-aminopropyl triethoxy silane (APTES), ethanol, acetic acid, glutaraldehyde, monoclonal anti-Mycobacterium tuberculosis LAM Clone CS-35 (NR-13811), purified lipoarabinomannan LAM H37Rv (NR-14848). The protocols for cleaning and activation followed by the functionalization of tapered fiber with APTES, glutaraldehyde and antibody monoclonal anti-Mycobacterium tuberculosis LAM were adapted from the related protocols reported in our earlier works and the other relevant literature [16–21].

2.2.2 Cleaning and activation treatment of tapered fibers

The tapered fibers undergo a thorough cleaning process to eliminate dust particles and debris that may lead to measurement errors. They are immersed in acetone for 10 minutes and subsequently air-dried. The activation of the cleaned surface is accomplished using a piranha solution. A fresh piranha solution is prepared by pipetting out 70 mL of concentrated sulfuric acid and mixing it with 30 mL of hydrogen peroxide. The fibers are carefully placed in a petri dish, and the piranha solution is gently poured over them until fully covered. The tapered fiber is left in the acidic solution for one hour, followed by meticulous washing with deionized water until the water reaches a neutral pH.

2.2.3 Amino functionalization on tapered fibers

The first step in the functionalization process involves creating amino groups on the surface of the tapered fiber. This procedure includes preparing a 1% APTES solution, utilizing ethanol and acetic acid as a solvent in a volume ratio of 5:2. In a falcon tube, 49.5 mL of the prepared solvent is combined with 0.5 mL of 3-aminopropyl triethoxy silane. The solution is thoroughly mixed on a vortex machine to ensure homogeneity. Subsequently, the activated tapered fibers are immersed in petri dishes containing the 1% APTES solution. The fibers are left in the solution for 10 minutes and then washed with ethanol. Finally, the functionalized fibers are positioned in an oven at 100°C for 1 hour.

2.2.4 Aldehyde functionalization on tapered fibers

During this functionalization step, aldehyde functional groups are introduced onto the surface of the tapered fiber. To achieve this, a 1% aqueous glutaraldehyde solution is prepared using deionized water. The initial 50% glutaraldehyde solution is diluted to 1% by pipetting 1 mL glutaraldehyde and combining it with 49 mL deionized water. The mixture is homogenized by shaking on a vortex machine for a few minutes. Next, the APTES-functionalized fibers are immersed in the glutaraldehyde solution for 20 minutes at room temperature. Finally, the fibers are washed with deionized water to complete the process.

2.2.5 Antibody immobilization on tapered fibers

Finally, the antibody is immobilized on the tapered fibers. The selected antibody, serving as a counterpart to the LAM (lipoarabinomannan) antigen, is the monoclonal anti-Mycobacterium tuberculosis LAM, Clone CS-35 (NR-13811), provided by BEI resources. The tapered optical fibers are immersed in the antibody solution (150 mg/mL) overnight at 4°C. The conjugation mediates the immobilization of the antibody on the surface between the amino group present in the anti-LAM CS-35 antibody and the aldehyde groups on the surface of the tapered fiber. Subsequently, BSA blocking is performed by submerging the optical fiber in a 0.5% BSA solution for 15 minutes at room temperature. This step with BSA helps prevent any nonspecific binding of biomolecules on the substrate surface.

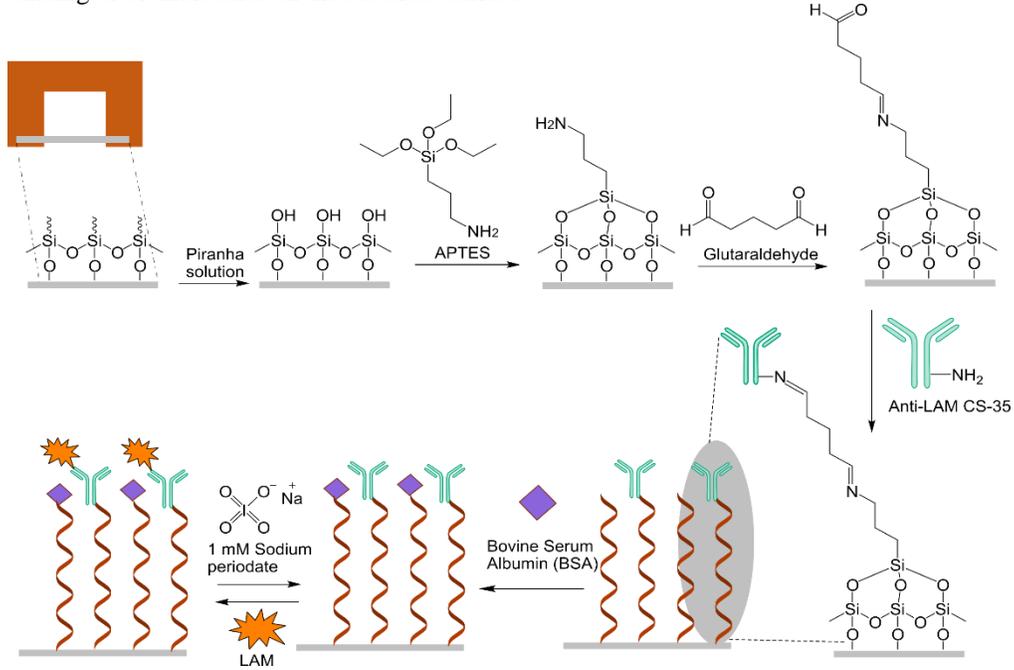

Figure 1: The schematic illustrates the surface functionalization process of a tapered fiber, comprising four key steps: amino functionalization, aldehyde functionalization, immobilization of the CS-35 antibody, and application of BSA blockers to reduce nonspecific binding.

## 2.3 Measurement Principle and Experimental Setup

We employ the principle of phase shift-cavity ring down spectroscopy (PS-CRDS), which involves measuring the phase shift, $\varphi$, of a sinusoidally modulated laser at the cavity output [22,23]. The phase shift, $\varphi$, is measured with respect to the sinusoid phase of the reference modulating signal and is related to the cavity photon lifetime, $\tau$, via the following equation:

$$\tan \varphi = \omega \tau, \qquad (1)$$

where $\omega$ is the amplitude modulation frequency. Fig. 2 shows the schematic diagram of our sensor that uses a fiber cavity for PS-CRDS measurements. The detailed experimental setup and measurement procedure are explained in our earlier work [15]. Briefly, we construct an optical fiber cavity of length 38 cm by splicing functionalized tapered fiber of waist diameter ≈ 10.2 µm in between two fiber Bragg gratings (FBGs). We place the functionalized tapered fiber assembly into a home-built fluidic cell to probe the sample effectively. We use an external Mach-Zehnder modulator to amplitude modulate a 1550 nm continuous wave laser diode with a 4 MHz and 2 $V_{pp}$ sinusoidal signal. Simultaneously, we current-modulate the source laser using a triangular wave (10 Hz, 50 mVpp) provided by a function generator. The PS-CRDS measurements at the cavity resonances are then recorded using a photodetector and a lock-in amplifier. The PS-CRDS measurements provide a phase shift, $\varphi$, with respect to a reference amplitude modulating signal. As the LAM binding on the functionalized tapered fiber leads to increased cavity losses, the ring-down time, $\tau$, decreases, consequently causing a reduction in the phase shift, $\varphi$, as predicted by Eq. (1).

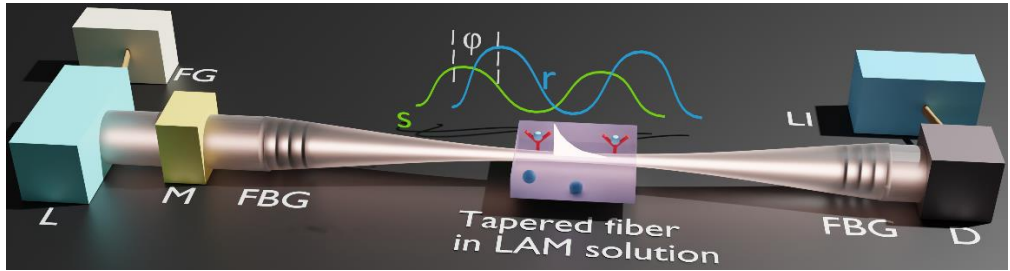

Figure 2: Experimental setup. L-Laser, FG-Function generator, M- Modulator, FGB-Fiber Bragg grating, LI: Lock-in amplifier, D-Detector. 's' represents the signal at the detector, and 'r' denotes the reference modulating signal from FG. The signal 's' undergoes a phase shift, φ, with respect to the reference signal 'r' due to LAM binding events on the functionalized tapered fiber.

## 3. Results

We conduct experiments at three stages to evaluate the sensor's performance. In the first stage, we test a bare tapered fiber, i.e., a fiber without any surface functionalization. We first immerse the tapered fiber in DI water and record the average phase shift values, $\varphi_w$, from 50–55 data points of absolute phase shift measurements. Then, we submerge the tapered fiber in a LAM solution (starting from low concentrations) and measure the phase shift value, $\varphi_s$. Our measurand, $\Delta\varphi = \varphi_s - \varphi_w$, denotes the difference between the phase shifts of LAM solution and DI water, as represented by red data points in Fig. 3. We repeat the same procedure to record the difference in phase shifts for different LAM concentrations as denoted by red data points in Fig. 3.

In the second stage, we use a tapered fiber functionalized with glutaraldehydes, step 4 of Fig. 1. We first immerse the tapered fiber in DI water to record the reference phase shift. We then immerse the fiber in the LAM solution for 15 minutes to ensure maximum LAM binding and measure the phase shift. After that, we treat the fiber with a 1 mM solution of sodium periodate for 15 minutes to detach the LAM molecules from the fiber. We then repeat this procedure for different LAM concentrations, and the results are shown by blue data points in Fig. 3.

In the third stage, we use a fully functionalized tapered fiber with LAM antibodies (see Fig. 1). We repeat the same measurement procedure as in the second stage, and the results are shown with black data points in Fig. 3.

The results show that the non-functionalized and tapered fiber coated with glutaraldehyde produces a very low response compared to LAM antibodies functionalized fiber, which confirms that the surface functionalization protocol and sensing measurements are working correctly. We also find that the sensor's LAM detection limit is 10 ppt (1 ppt= 1 pg/mL), and from the linear fit of black data points, we can infer that the sensor's sensitivity is 0.026°/ppt.

## 4. Conclusions

Our work showcases the pioneering use of a phase shift cavity ringdown spectroscopy in conjunction with fiber cavities toward rapid TB diagnostics. Our experiments demonstrate that we can detect LAM as a TB biomarker at concentrations as low as 10 pg/mL, which provides almost fifty times improvement compared to the previous work [14]. Compared to conventional and earlier LAM sensors, our demonstrated sensor method offers significant advantages, including higher sensitivities, quicker detection, reduced costs, minimal sample preparation requirements, simplified system complexity, fewer fabrication difficulties, and enhanced potential for portability.

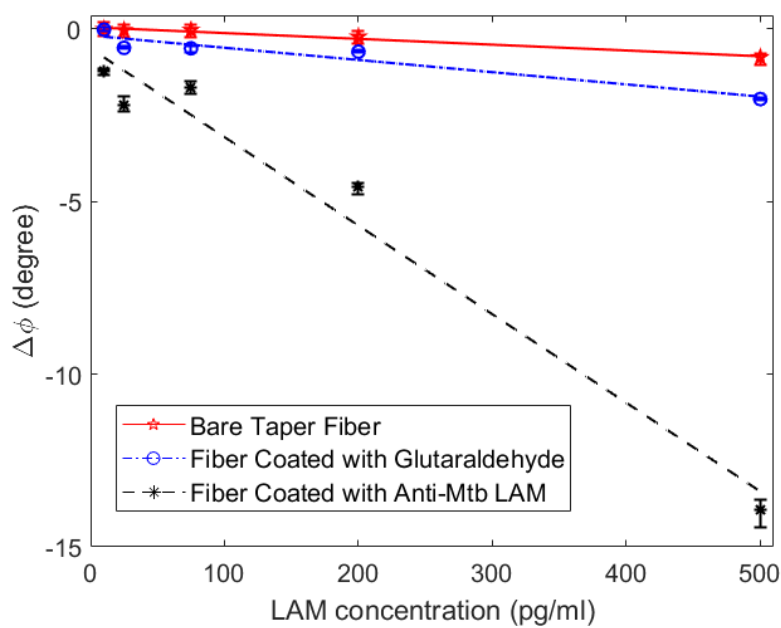

Figure 3: The experimental results depict the sensor response for three types of tapered fibers. The slope of the line is highest for the fiber functionalized with the CS-35 antibody, while the fiber without any functionalization produces a minimal response. This successfully demonstrates the correct fiber chemical functionalization and accurate functioning of the sensor.

There are several potential avenues for enhancing this research in the future. These avenues encompass conducting tests with urine samples, implementing temperature control for the fiber Bragg gratings (FBGs) and the fluidic cell, and refining and optimizing various parameters for tapering the fibers to increase the sensor's sensitivity. We foresee that the sensor setup we have demonstrated, with modifications to the surface chemistry protocol, could evolve into an affordable, real-time, and dependable optical sensor applicable across a wide range of sectors, such as food safety, healthcare, environmental monitoring, and agriculture, especially in resource-constrained environments.

**Funding.** CureMD grant#G109

**Disclosures.** MF: CureMD Healthcare, INC, USA (F)

**Data availability.** Data underlying the results presented in this paper are not publicly available at this time but may be obtained from the authors upon reasonable request.

**References**

1. S. Bagcchi, "WHO's Global Tuberculosis Report 2022," The Lancet Microbe **4**(1), e20 (2023).
2. B. Acharya, A. Acharya, S. Gautam, S. P. Ghimire, G. Mishra, N. Parajuli, and B. Sapkota, "Advances in diagnosis of Tuberculosis: an update into molecular diagnosis of Mycobacterium tuberculosis," Mol Biol Rep **47**(5), 4065–4075 (2020).
3. A. L. García-Basteiro, A. DiNardo, B. Saavedra, D. R. Silva, D. Palmero, M. Gegia, G. B. Migliori, R. Duarte, E. Mambuque, R. Centis, L. E. Cuevas, S. Izco, and G. Theron, "Point of care diagnostics for tuberculosis," Pulmonology **24**(2), 73–85 (2018).
4. M. Phillips, V. Basa-Dalay, J. Blais, G. Bothamley, A. Chaturvedi, K. D. Modi, M. Pandya, M. P. R. Natividad, U. Patel, N. N. Ramraje, P. Schmitt, and Z. F. Udwadia, "Point-of-care breath test for biomarkers of active pulmonary tuberculosis," Tuberculosis **92**(4), 314–320 (2012).
5. S. Gupta and V. Kakkar, "Recent technological advancements in tuberculosis diagnostics – A review," Biosensors and Bioelectronics **115**, 14–29 (2018).

6. S. H. Chauke, S. Nzuza, S. Ombinda-Lemboumba, H. Abrahamse, F. S. Dube, and P. Mthunzi-Kufa, "Advances in the detection and diagnosis of Tuberculosis using optical-based devices," Photodiagnosis and Photodynamic Therapy 103906 (2023).
7. C. Maphanga, S. Manoto, S. Ombinda-Lemboumba, Y. Ismail, and P. Mthunzi-Kufa, "Localized surface plasmon resonance biosensing of Mycobacterium tuberculosis biomarker for TB diagnosis," Sensing and Bio-Sensing Research **39**, 100545 (2023).
8. B. A. Prabowo, Y.-F. Chang, H.-C. Lai, A. Alom, P. Pal, Y.-Y. Lee, N.-F. Chiu, K. Hatanaka, L.-C. Su, and K.-C. Liu, "Rapid screening of Mycobacterium tuberculosis complex (MTBC) in clinical samples by a modular portable biosensor," Sensors and Actuators B: Chemical **254**, 742–748 (2018).
9. J. M. Hong, H. Lee, N. V. Menon, C. T. Lim, L. P. Lee, and C. W. M. Ong, "Point-of-care diagnostic tests for tuberculosis disease," Sci Transl Med **14**(639), eabj4124 (2022).
10. U. Ullah, Z. Tahir, O. Qazi, S. Mirza, and M. I. Cheema, "Raman spectroscopy and machine learning-based optical probe for tuberculosis diagnosis via sputum," Tuberculosis (Edinb) **136**, 102251 (2022).
11. B. Hamasur, J. Bruchfeld, M. Haile, A. Pawlowski, B. Bjorvatn, G. Källenius, and S. B. Svenson, "Rapid diagnosis of tuberculosis by detection of mycobacterial lipoarabinomannan in urine," J Microbiol Methods **45**(1), 41–52 (2001).
12. J. Flores, J. C. Cancino, and L. Chavez-Galan, "Lipoarabinomannan as a Point-of-Care Assay for Diagnosis of Tuberculosis: How Far Are We to Use It?," Front Microbiol **12**, 638047 (2021).
13. M. A. Bulterys, B. Wagner, M. Redard-Jacot, A. Suresh, N. R. Pollock, E. Moreau, C. M. Denkinger, P. K. Drain, and T. Broger, "Point-Of-Care Urine LAM Tests for Tuberculosis Diagnosis: A Status Update," J Clin Med **9**(1), 111 (2019).
14. P. Ramirez-Priego, D. Martens, A. A. Elamin, P. Soetaert, W. Van Roy, R. Vos, B. Anton, R. Bockstaele, H. Becker, M. Singh, P. Bienstman, and L. M. Lechuga, "Label-Free and Real-Time Detection of Tuberculosis in Human Urine Samples Using a Nanophotonic Point-of-Care Platform," ACS Sens. **3**(10), 2079–2086 (2018).
15. M. D. Ghauri, S. Z. Hussain, U. Ullah, R. M. A. Ayaz, R. S. Z. Saleem, A. Kiraz, and M. I. Cheema, "Detection of aflatoxin M1 by fiber cavity attenuated phase shift spectroscopy," Opt. Express, OE **29**(3), 3873–3881 (2021).
16. S. Nayab, A. Farrukh, Z. Oluz, E. Tuncel, S. R. Tariq, H. ur Rahman, K. Kirchhoff, H. Duran, and B. Yameen, "Design and Fabrication of Branched Polyamine Functionalized Mesoporous Silica: An Efficient Absorbent for Water Remediation," ACS Appl. Mater. Interfaces **6**(6), 4408–4417 (2014).
17. A. Butt, A. Farrukh, A. Ghaffar, H. Duran, Z. Oluz, H. ur Rehman, T. Hussain, R. Ahmad, A. Tahir, and B. Yameen, "Design of enzyme-immobilized polymer brush-grafted magnetic nanoparticles for efficient nematicidal activity," RSC Adv. **5**(95), 77682–77688 (2015).
18. T. D. To, A. T. Nguyen, K. N. T. Phan, A. T. T. Truong, T. C. D. Doan, and C. M. Dang, "Modification of silicon nitride surfaces with GOPES and APTES for antibody immobilization: computational and experimental studies," Adv. Nat. Sci: Nanosci. Nanotechnol. **6**(4), 045006 (2015).
19. A. Bekmurzayeva, Z. Ashikbayeva, N. Assylbekova, Z. Myrkhiyeva, A. Dauletova, T. Ayupova, M. Shaimerdenova, and D. Tosi, "Ultra-wide, attomolar-level limit detection of CD44 biomarker with a silanized optical fiber biosensor," Biosensors and Bioelectronics **208**, 114217 (2022).
20. Y. J. Chuah, S. Kuddannaya, M. H. A. Lee, Y. Zhang, and Y. Kang, "The effects of poly(dimethylsiloxane) surface silanization on the mesenchymal stem cell fate," Biomater. Sci. **3**(2), 383–390 (2015).
21. G. Rashidova, M. Tilegen, T. T. Pham, A. Bekmurzayeva, and D. Tosi, "Functionalized optical fiber ball-shaped biosensor for label-free, low-limit detection of IL-8 protein," Biomed. Opt. Express, BOE **15**(1), 185–198 (2024).
22. R. Engeln, G. von Helden, G. Berden, and G. Meijer, "Phase shift cavity ring down absorption spectroscopy," Chemical Physics Letters **262**(1), 105–109 (1996).
23. M. I. Cheema, S. Mehrabani, A. A. Hayat, Y.-A. Peter, A. M. Armani, and A. G. Kirk, "Simultaneous measurement of quality factor and wavelength shift by phase shift microcavity ring down spectroscopy," Optics express **20**(8), 9090–9098 (2012).